# A Feasibility Study on Real-Time High Resolution Imaging of the Brain Using Electrical Impedance Tomography

**Sébastien Martin[a1]**

[a] Department of Electrical and Computer Engineering, National Chiao Tung University
[1] sebastien1606.eed01g@nctu.edu.tw

## Abstract

*Objective:* The strengths of Electrical Impedance Tomography (EIT) are its capability of imaging the internal body by using a noninvasive, radiation safe technique, and the absence of known hazards. In this paper we introduce a novel idea of using EIT in microelectrodes during Deep Brain Stimulation (DBS) surgery in order to obtain an image of the electrical conductivities of the brain tissues surrounding the microelectrodes. DBS is a surgical treatment involving the implantation of a medical probe inside the brain. For such application, the EIT reconstruction method has to offer both high quality and robustness against noise. *Methods:* A post-processing method for open-domain EIT is introduced in this paper, which combines linear and nonlinear methods in order to use the advantages of both with limited drawbacks. The reconstruction method is a two-steps method, the first solves the inverse problem with a linear algorithm and the second brings the nonlinear aspects back into the image. *Results:* The proposed method is tested on both simulation and phantom data, and compared to three widely used method for EIT imaging. Resulting images and errors gives a strong advantage to the proposed solution. *Conclusion:* High quality reconstruction from phantom data validates the efficiency of the novel reconstruction method. *Significance:* This feasibility study presents an efficient method for open domain EIT and opens the way to clinical trials.

**Highlights:**

A real-time two-dimensional imaging of the inner part of the brain
Imaging the inner activity of the brain during deep brain stimulation surgery
Simulation of open-domain imaging of the sub thalamic nucleus region
A high resolution biomedical imaging method using based on electrical properties

# 1 INTRODUCTION

Electrical Impedance Tomography (EIT) is a modern technique proposed for low-cost and patient's safe biomedical imaging. The basic idea to use a pair of electrodes to send an electrical current through a volume conductor and to measure the voltages with other electrodes located all around this volume. By applying a certain electric current at the boundaries of the volume, it is possible to obtain the internal distribution of electrical conductivities by measuring the electrical voltages and solving an inverse problem (Holder, 2005).

EIT is usually used to measure the conductivities of internal tissues by placing electrodes around the human body. It has already given good results in producing images of the lungs (Kuen et al., 2009) and can be used for reconstructions of other parts of the body (Brown, 2003). In this paper, EIT technique is used for the first time to obtain a map of the electrical conductivities of the brain tissues surrounding microelectrodes used during a Deep Brain Stimulation (DBS) surgery (Perlmutter and Mink, 2006). The goal of such mapping is to create a real-time and accurate map of the area surrounding the microelectrodes.

DBS is a surgical technique which aims to send an electrical signal directly to a specific area of the brain to reduce disabling neurological symptoms (for instance, a tremor) (Liker et al., 2008). The idea of using an artificial electrical current for brain healing is quite old, but the knowledge and the equipment necessary to actually perform DBS on human brains are relatively new (McIntyre et al., 2015). Currently it is seen as a promising technique to heal many brain disorders, such as Parkinson's disease (Lozano, 2001), epilepsy (Boon et al., 2007), and Alzheimer's disease (Laxton et al., 2010). However, one of the major difficulties of DBS is to insert permanent leads (or electrodes) at the precise location in the brain to impact on the symptoms.

To locate the target area, a Magnetic Resonance Imaging (MRI) (Vayssiere et al., 2000) or Computed Tomography (CT) (Fiegele et al., 2008) scan is performed prior to surgery. These give a high quality, three-dimensional map of the different regions of the brain, which later helps to pinpoint the target area. However, this scan itself is usually not sufficient and the surgery relies to some extent on trial and error to obtain a precise position. During this trial and error phase, inserted microelectrodes can be used to measure the neuronal activity of the surrounding tissue. This stage of the surgery is time-consuming and requires the cooperation of the patient. Since the brain is naturally using electrical signals to stimulate the neurons, applying 'artificial' electrical stimulation at this stage may quickly lead to side effects such as degradation of the brain (Baizabal Carvallo et al., 2012). While using microelectrodes to emit a very low current below the threshold is not harmful for the brain, it is highly inadvisable to use a high current during the trial and error stage.

The final goal of this paper is to demonstrate the possibility to use EIT as a mapping tool that is capable of reconstructing an accurate map of the area of the brain surrounding the microelectrode. These will retrieve information about living tissues located only a few millimeters from the microelectrode. This is achieved by using a modified version of a microelectrode to send an extremely low electrical current, and by solving the EIT inverse problem to obtain an image of the conductivity of the different brain tissues surrounding the probe. Nonlinear algorithms, such as an Artificial Neural Network (ANN), can be applied to the resulting image, as a post processing tool (Martin and Choi, 2017), in order to refine the quality of the reconstruction.

# 2 MATERIAL AND METHODS

## 2.1 Electrical Impedance Tomography

EIT is a modern method of tomography. By sending an electrical signal into electrically conductive tissues, it is possible to obtain the electrical properties of the living tissues that have been crossed by this current. Assuming that different tissues have different properties from each other's, it becomes possible to use the electrical measurements to generate a map of the tissues within the scanned area. However, solving the inverse problem (i.e. given the measurement at the boundary, recovering the original electrical conductivity of the different elements of the FE model) is not an easy task (Leonhardt and Lachmann, 2012). This inverse problem is severely ill-posed, and a common practice is to assume the solution to be close to an initial guess and to search for a linear variation of this initial estimate. This implies that the electrical conductivity of one element also depends on the conductivity of its neighbors. Mathematically, this is done by using a linear reconstruction algorithm and prior probability functions in an inverse solver (Bayford, 2006).

EIT simulations can be divided into two sub-problems, respectively termed the forward problem and the inverse problem.

### 2.1.1 The forward problem

In the forward problem, the objective is to obtain the resulting voltages at the boundaries, given the input current and the conductivities of the different elements in the FE model. In EIT simulations, this problem has to be solved first, to obtain the voltage distribution at the boundary, and can be solved by

Maxwell's equation on electromagnetism (Cheney et al., 1999):

$$\nabla \cdot \sigma(x,\omega)\nabla u = 0 \quad (1)$$

where, for each point $x$ and each angular frequency $\omega$, $\sigma$ is the electric conductivity, and $u$ is the electric potential in a body $\Omega$. Given the forced boundary condition (Xu et al., 2005):

$$u = u_0 \quad (2)$$

where $u_0$ is the boundary potential, and the Neumann's boundary condition:

$$\sigma \frac{\partial u}{\partial n} = -J \quad \text{on } \partial\Omega \quad (3)$$

where n is the unit outward normal vector at the boundary, and J is the current density at the boundary. After discretization in the FE model, this linear system of equations can be solved in order to obtain the distribution of the electrical current $J$ at the boundaries of the system.

### 2.1.2 The inverse problem

Although solving the forward problem can be done linearly, the inverse problem is highly non-linear and more complicated to solve. Since the inverse problem is severely ill-posed, solving it requires either simplification of the mathematical approach by assuming an initial guess of the solution and considering the actual solution to be a linear perturbation from this guess, or using a non-linear optimization method which can approximate the solution.

On the other hand, it has been proven that non-linear algorithms can find an optimal solution to the EIT problem, without extensively relying on the initial guess. These algorithms, usually based on Evolutionary Algorithm (EA) (Zhou et al., 2007) or ANN (Wang et al., 2009), are known to be capable of approximating the solution of a non-linear problem. By applying them to solve the EIT inverse problem, it becomes possible to obtain an estimate of distribution of the electric admittivity of the different elements within the FE model.

### 2.1.3 Open-domain EIT

A major concern in open domain EIT is the density of the mesh. In bounded domain EIT, it has been demonstrated that reconstructing an object located at the center of the tank usually gives more distortion than reconstructing an object located close to the electrodes, at the boundary of the tank.

Fig 1 illustrates the idea of open-domain EIT. In the case of unbounded domain EIT, this situation may depend on the inverse solver and prior probability used. Though, usually, when the object is located far from the probe, high distortion may appear on the resulting image. This is due to the rapid decay of sensitivity when increasing the distance from the probe (A Borsic et al., 2010). Also, it is important to note that solving an inverse problem in an open domain may lead to high reverberation effects at the boundaries (Chen and Konrad, 1997). Reverberations at the outer boundary create additional distortion in the FE model when solving the inverse problem.

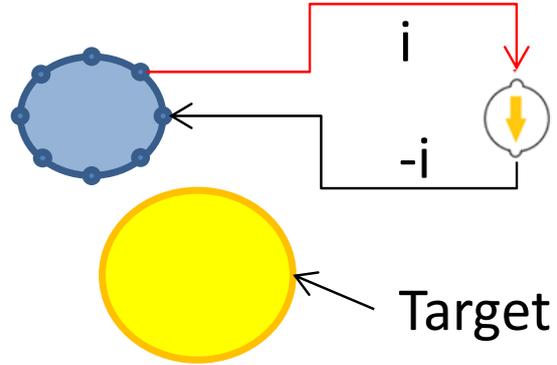

Fig 1 Illustration of the open-domain EIT problem. The DBS probe, having 8 electrodes, is shown in blue and the target is shown in yellow

To avoid such reverberation phenomenon, a common practice is to use an excessively large mesh, which has an outer boundary farther than the physical phantom. Several methods have been discussed to determine the size and the limit of the outer boundary (Chen and Konrad, 1997). It appears that a widely used method is to choose a mesh at least five times larger than the distance between the most remote outer target object and the center of the probe. Since such a large mesh will require a very intense computation to generate the solution to the inverse problem, the density of the mesh increases gradually the further we move from the probe.

## 2.2 Deep Brain Stimulation

Deep Brain Stimulation is an advanced neurosurgical application which uses electrical stimulation to modulate the activities in a specific target area of the brain to reduce the disabling neurological symptoms in the patients. To do that, an electrical signal is transmitted to a specific area of the brain, aiming to activate some specific neurons. Although the idea was first conceived in the 50's (McIlwain, 1951), there has been an increase in DBS applications only in the last twenty years. Despite this increase, this promising technology still faces some difficulties (Chaturvedi, 2012).

One of the main difficulties of DBS is to accurately locate the target area within the brain. An MRI scan is usually performed prior to the insertion of a DBS probe. However, such a scan can only produce an image of the brain prior to the insertion of the probe. The insertion of a probe may lead to displacement of the brain tissue (Winkler, 2005). Also, previous studies have noted that, with the incorrect placement of a DBS probe, the surgery may lead to undesirable side-effects, including apathy, hallucinations, compulsive gambling, hyper sexuality, cognitive dysfunction, and depression (Burn and Tröster, 2004). Therefore, in order to make sure that the electric stimulation will only stimulate the desired target, it would be highly desirable to obtain both real time measurement and a real time image reconstruction of the surrounding area.

The basic difference between EIT and DBS is in the intensity of stimulation. In EIT, the electrical current is kept as low as possible, so that it does not stimulate the neurons and start a chain reaction, as DBS does. Further, using a low current does not produce the undesirable side effects of DBS noted above.

In this paper, we explore the concept of using the DBS probe, which contains electrodes capable of both emitting and receiving an electrical current, to map the electrical distribution of the tissues surrounding the probe. From this electrical distribution, it then becomes possible to recognize different shapes with a very high degree of accuracy. Based on CT or MRI images obtained prior to surgery, it becomes possible to follow the development of the shape and location of the target after inserting the probe, while increasing the quality of the images.

## 2.3  Artificial Neural Networks

Artificial Neural Networks are a series of different algorithms that can approximate any kind of nonlinear function (Leshno et al., 1993), and these powerful Artificial Intelligence (AI) algorithms can approximate a solution to the nonlinear problem of post processing.

An ANN attempts to imitate the behavior of the human brain, or a natural neural network, and can give a nonlinear response to an external stimulation. A neural network (natural or artificial) is up made of simple elements, called neurons, which are interconnected. Each neuron can be modeled by a linear mathematical model. An input 'signal' stimulates the neuron, which produces a response, depending on the input signal. In an ANN, where all neurons are interconnected in complex ways, the output of one neuron can become the input of another neuron (Hu Hen and Hwang, 2001).

An ANN is usually made of three different layers of neurons. The first layer, the input layer, is generally used to weight and transmit the input value to the second layer. The second layer of neurons then computes a specific output, depending on the input values, weights and biases. Then, the result is sent to the third layer of neurons, called the output layer. Between two different layers, the signal is adjusted with biases and weights. These weights and biases are fixed parameters that are determined prior to using the ANN, during the training phase. During this phase, the network is fed with input data and output values are computed. These output values are then compared to the correct values (in this application, the simulated conductivity distributions). By comparing the output of the ANN and the target, it is possible to compute an error and to use it to adjust the weights and biases via a specific algorithm called a training algorithm. Different training algorithms have previously been proposed (Hu Hen and Hwang, 2001), giving different results depending on the original problem. For EIT inverse problem, the particle swarm optimization method is known to be effective (Martin and Choi, 2016).

However, it should be noted that it is difficult for an ANN to generalize a problem when the number of input or output data become too large. This is a consequence of the increasing number of neurons, weights and biases. Since the size of the original FE model is significantly large to avoid distortion when solving the inverse problem using a linear algorithm, it is better here to avoid having too large a number of elements.

Principal Component Analysis (PCA) is a commonly used method to reduce the number of inputs of a neural network. The basic idea of PCA is to reduce the number of inputs by selecting only the input data with the most important features. In this paper, the size of the mesh has been reduced significantly before applying ANN. Since the object is supposed to be located within a distance of 5 times the radius of the probe, the nodes located farther than this distance are only useful for avoiding reverberation effects when solving the inverse problem, and the conductivity of the nodes located far from the probe then have the conductivity of the background (or contain distortion). By assuming that the most relevant information is contained at the nodes located close to the probe, it is possible to feed the neural network with those nodes only, without losing any significant information.

In this paper, an ANN is used to improve the EIT images obtained by means of linear inverse solvers. The aim of post-processing is to use AI algorithm to add a nonlinear behavior in the solution of the inverse problem. By adding a nonlinear algorithm after applying a linear inverse solver algorithm, the global solution becomes a nonlinear problem. Since the EIT inverse problem is a nonlinear problem, the solution presented is therefore supposed to give more accurate image reconstruction.

## 2.4  Methods

This section describes how the experiments were carried out to generate results that can be reproduced by anyone.

In order to avoid any reverberation effects, the FE model was significantly larger than the maximum distance from the probe to an object. While the centroid of the object is not more than 10 mm from the boundary of the probe, the external boundary of the open-domain EIT tank is 70 mm from the center of the tank. A circular DBS probe is simulated at the center by placing 8 electrodes around a circle.

To train the ANN, a training data set of 2 000 EIT images was generated. The presence of an elliptical object of a reasonable size and conductivity was simulated in the FE model, close to the probe. The centers of gravity of the targets were located anywhere within a radius of 10 times the radius of the probe. Images were obtained using the FE method with a large mesh of 5007 nodes and 9946 elements. Reconstructions algorithms used EIDORS toolkit (Adler and Lionheart, 2006) and MATLAB's ANN toolbox. CPU times were measured on an Intel Core I7 6700k CPU running Ubuntu Linux. The training was performed using the Particle Swarm optimization method (Martin and Choi, 2016).

### 2.4.1 Data acquisition

Phantom data were acquired using the data acquisition system developed in (Tu, 2015). This system uses pairs of adjacent electrodes for current injection and voltage measurements. For each current injection, 8 different voltages were measured with the 8 electrodes present at the boundary of the probe. The current source was then moved to the next pair of electrodes and 8 further measurements were made. Finally, a total of 64 measurements were used for image reconstruction. Although this method only gave 32 independent measurements, it is common practice to keep all the measurements in order to reduce the influence of noise (Miao et al., 2014). Measurements obtained with electrodes used for current injection were not kept for the reconstruction. The probe was inserted into a large water tank in order to avoid reverberation effects and minimize the influence of the physical boundary. A photo of the water tank containing the experimental probe is shown in Fig 2.

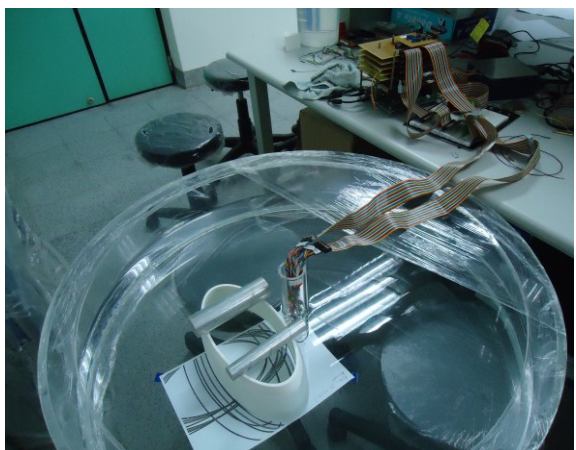

Fig 2 The experimental probe submerged in ionized water in the middle of the water tank

### 2.4.2 Noise estimate

The presence of noise was taken into consideration during phantom experiments. Although most of the noise was cancelled by using a tenth-order bandpass filter centered on the frequency of the injected current, some noise was still present in the data acquired. In order to reduce the need to extrapolate using ANNs, it was expedient to train them with noisy data similar to the data acquired from the phantom.

The amount of noise is not fixed and strongly depends on the physical distance between the current source and the electrodes used for measurement, as well as the medium the current has to cross. Noise was estimated independently for all measured data. Current injection consisted of a sinusoidal waveform at 100 kHz. During each sine wave, 20 voltages were measured by each pair of electrodes. For EIT reconstruction, these data were filtered and the highest peak was used. The noise was estimated by comparing the measured data to a simulated sine wave. By estimating the amount of noise before filtering and assuming the noise to be WGN, one can estimate the noise level for the different measurements. This procedure is given in detail in (Martin and Choi, 2016).

### 2.4.3 Error calculation

To obtain an objective definition of the image quality, an objective error function has to be defined, and several methods have been proposed for medical imaging applications (Abascal et al., 2008; Andrea Borsic et al., 2010). Among these definitions, the difference of Resolution (|ΔRES|) and Shape Deformation (SD) are particularly relevant to this application.

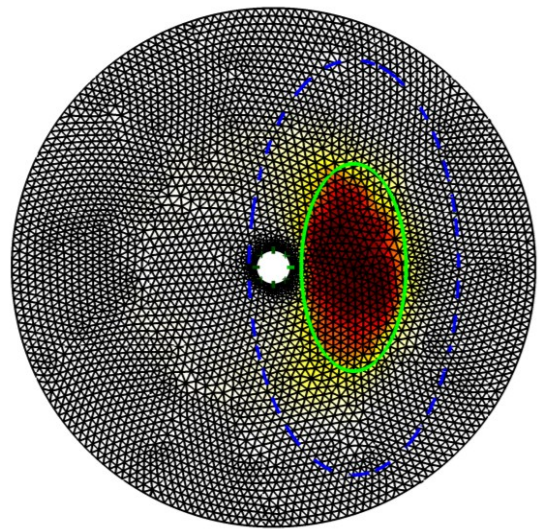

Fig 3 Illustration of ADE. An EIT image is reconstructed. The original target is shown by a green ellipse. The ROI, an ellipse with double the dimensions of the target, is shown by blue dashed line. The resulting image is converted to a binary image by thresholding and the ADE is computed on the thresholded image.

ADE considers a specific ROI around the target. This ROI has the same shape as the target, but is twice as large. The image threshold was similar to |ΔRES|, and was fixed at 25% of the maximal conductivity. ADE then compares the ROI of the binary image obtained from reconstruction after

applying the thresholding. Parts of the reconstructed image containing an object (typically, where a difference of conductivity is visible) where the initial image (the simulated target) does not, and vice versa, are considered as an error. The area of the error region is then computed, divided by the perimeter of the elliptical target and normalized to the diameter of the probe. ADE is computed by (4):

$$ADE = \frac{A_{error}}{P_{target}} \quad (4)$$

where $A_{error}$ is the area of the error region within the ROI, $P_{target}$ is the perimeter of the target. ADE is then normalized to the diameter of the probe to get a Normalized ADE (NADE).

The Region Of Interest (ROI) used in ADE calculation is similar to the ROI used for calculating the GREIT-based errors |ΔRES| and SD. Here, the ROI is an ellipsoid centered on the ellipsoidal target object, as shown in Fig 3. The ROI, with radii twice as large as the radii of the target, is represented by a blue dotted line. For all different reconstructions, either from simulation or phantom experiments, the resulting errors are being compared based on the distance, edge to edge, between the probe and the target.

# 3 RESULTS
## 3.1 Scaling

This section aims to demonstrate that it is possible to use a larger probe for phantom experiments, as long as the dimensions of the target, as well as its distance from the probe, are proportional to the probe. For example, in a biomedical environment, a probe having a diameter of at least 1.5 mm will be used to image the STN region, a 4*6mm ellipsoid. In phantom experiments presented in this paper, a prototype probe having a diameter of 50 mm is being used, and the target region was designed proportionally. Simulation studies were conducted with three different probes having a diameter of 1.5 mm, 9 mm, and 50 mm respectively. For each case, the target was located at different distances from the probe (normalized based on the dimension of the probe), and its size was proportional to the size of the probe. Corresponding inverse problems were solved with the one-step GN, PDIPM, and the proposed post-processing. Following image reconstruction, |ΔRES|, NADE, and SD errors were computed and the results are shown in Table 1, Table 2, and Table 3, respectively.

It appears that the three different cases give a similar error. This observation means that it is possible to use a large prototype probe for phantom experiments, as long as the size and location of the target is scaled to the size of the probe.

Table 1 |ΔRES| errors obtained with three different probes for an elliptical target located at a distance of 1.5 times the diameter of the probe.

| Method \ ProbeØ | 1.5mm | 9mm | 50mm |
|---|---|---|---|
| One-step GN | 33.31% | 33.36% | 33.54% |
| PDIPM | 28.10% | 28.16% | 28.48% |
| One-step GN + ANN | 12.03% | 12.11% | 12.54% |

Table 2 NADE errors obtained with three different probes for an elliptical target located at a distance of 1.5 times the diameter of the probe.

| Method \ ProbeØ | 1.5mm | 9mm | 50mm |
|---|---|---|---|
| One-step GN | 4.67 | 4.62 | 4.62 |
| PDIPM | 3.45 | 3.41 | 3.43 |
| One-step GN + ANN | 1.04 | 1.04 | 1.03 |

Table 3 SD errors obtained with three different probes for an elliptical target located at a distance of 1.5 times the diameter of the probe.

| Method \ ProbeØ | 1.5mm | 9mm | 50mm |
|---|---|---|---|
| One-step GN | 61.20% | 61.31% | 61.59% |
| PDIPM | 56.35% | 56.34% | 56.68% |
| One-step GN + ANN | 39.01% | 39.43% | 41.08% |

## 3.2 Simulation

The proposed method was first tested with simulation data in order to validate the idea of post processing. For simulation, ANNs were trained without adding any noise in the simulated voltages. 40 different images, representing the STN region at different distances, were simulated and their corresponding inverse problems were solved by the 4 different methods.

Some of the images obtained are given in Fig 4, which shows the STN region located at five different distances from the probe. The distances between the probe and the STN region are, respectively, 1, 2, and 3 times the radius of the DBS probe.

As expected from the theory, the one-step GN solver generates large smoothness in the conductivity distribution, and the resulting image does not really show the location of the ROI. Although this method seems to be capable of giving a correct approximation of the direction of the target, it does not give a satisfactory estimate of the distance between the probe and the target ROI. As the target ROI becomes further from the probe, the one-step GN method outputs a linear conductivity distribution with a red ROI (which is assumed to be the target object) near the outer boundary of the domain. The target object cannot be seen at the expected location and therefore the resulting errors are high.

With the nonlinear iterative PDIPM method, the resulting conductivity distributions show a lower conductivity distributions show a lower conductivity within the region of interest, although a large smoothness remains. For instance, in Fig 4 (b), when the target is relatively close to the probe, it can be seen that the PDIPM solver successfully estimates a satisfactory estimate of half of the profile of the target object. The part of the target which is 'directly' visible from the probe can be imaged without difficulty. However, the other side of the target object, which is both far from the probe and not directly visible, is not satisfactorily shown, and a smooth conductivity distribution is seen instead. As the distance from the probe increases, the ringing becomes more intense and therefore the reconstructed object is less easily visible. However, it can be seen (Fig 4 (l)) that it is also difficult to give a satisfactory estimate of the complete profile of the STN region with this reconstruction algorithm.

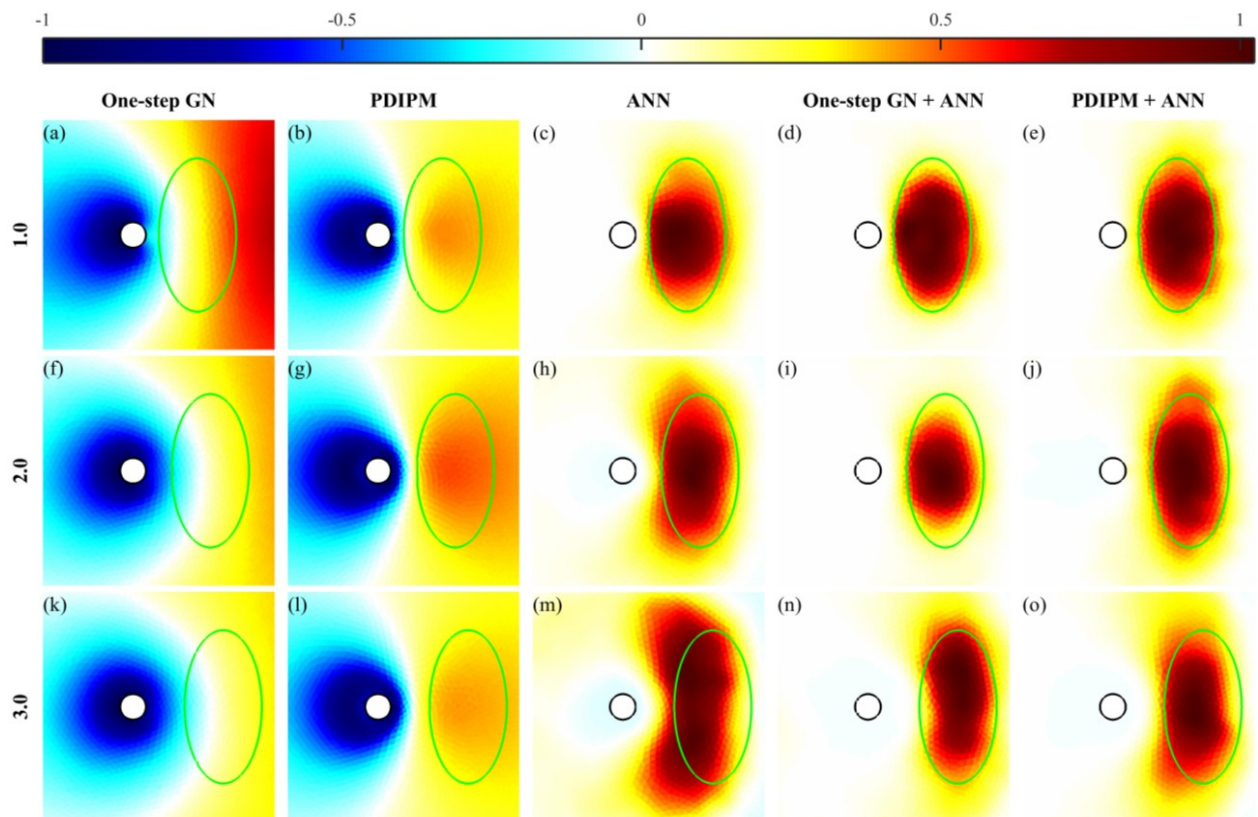

Fig 4 EIT image reconstruction obtained from simulation data with existing methods and the proposed one. Images were reconstructed with (1st column) one-step GN, (2nd column) PDIPM, (3rd column) ANN as inverse solver, (4th column) the proposed post-processing applied after the one-step GN, and (5th column) the proposed post-processing applied after PDIPM solver. The normalized distances between probe and target are 1, 2, and 3, respectively on 1st, 2nd, and 3rd rows. The probe is shown by a black circle and the target ROI by a green ellipse. On top, the normalized resistiviy distribution.

When simulation data are being used, an ANN used as an inverse solver can generate a very high quality reconstruction, since the input data are close to the data used for training the ANN and the need of extrapolation is therefore minimal. It can also be seen in Fig 4 that an ANN used as an inverse solver can give a satisfactory estimate of the profile of the STN region, near the probe as well as near the outer boundary. An accurate estimate of the complete contour of the target was difficult to obtain with the iterative PDIPM method. An accurate estimate of the profile of the target object led to a more accurate estimate of the size of the target as well as smaller errors.

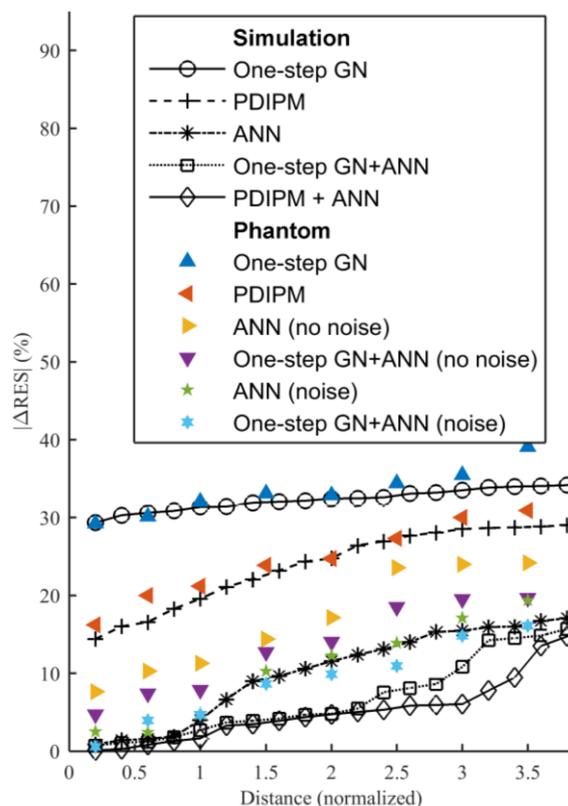

Fig 5 |ΔRES| errors obtained for simulation and phantom data. Errors are given for different reconstruction methods for a target located at a variable distances from the probe.

Finally, using an ANN as a post-processing tool also offers high quality reconstruction. Fig 4 (4th column) shows that this method gives an accurate estimate of the size and location of the STN. The reconstructed target fits within the green ellipse that represents the location of the original target. Similar to the ANN used as an inverse solver, the proposed method generates rough conductivity changes around the target boundary. Despite a loss of information induced by the linear one-step GN method, it can be seen that applying an ANN as a post-processing tool still gives an accurate estimate of the nonlinear conductivity distribution at a distance of up to three times larger than the radius of the probe.

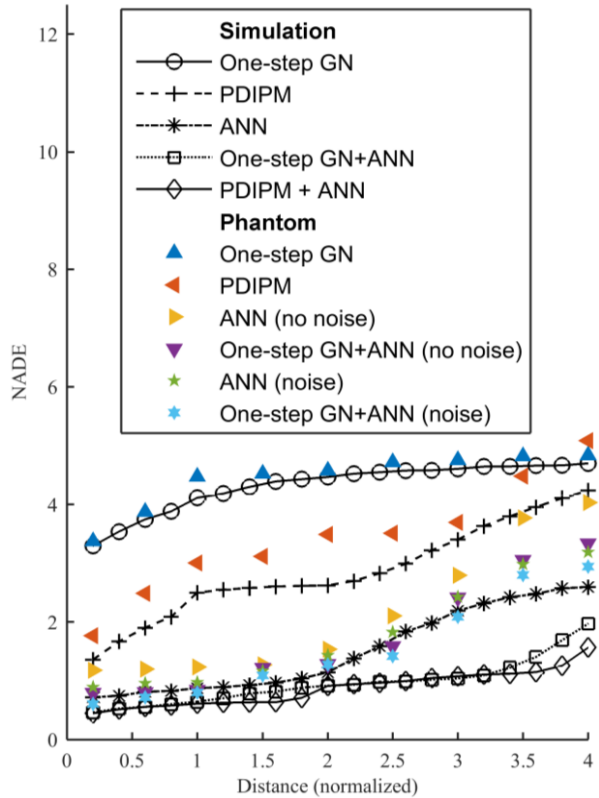

Fig 6 NADE obtained for simulation and phantom data. Errors are given for different reconstruction methods for a target located at a variable distances from the probe.

|ΔRES| errors presented in Fig 5 show that the proposed post processing gives better results than the ANN used as a direct inverse solver when the target object is located at a distance equivalent to the radius of the probe. An ANN used as a direct inverse solver gives an error comprised between 0.6% and 17.3%, which is significantly better than the one-step GN (from 29.2% to 34.7%) or PDIPM (from 14.1% to 29.7%). Finally, the proposed post-processing method gives a |ΔRES| error comprised between 0.1% and 16.2% The proposed method has consistently outperformed all other methods, including the ANN as inverse solver, significantly.

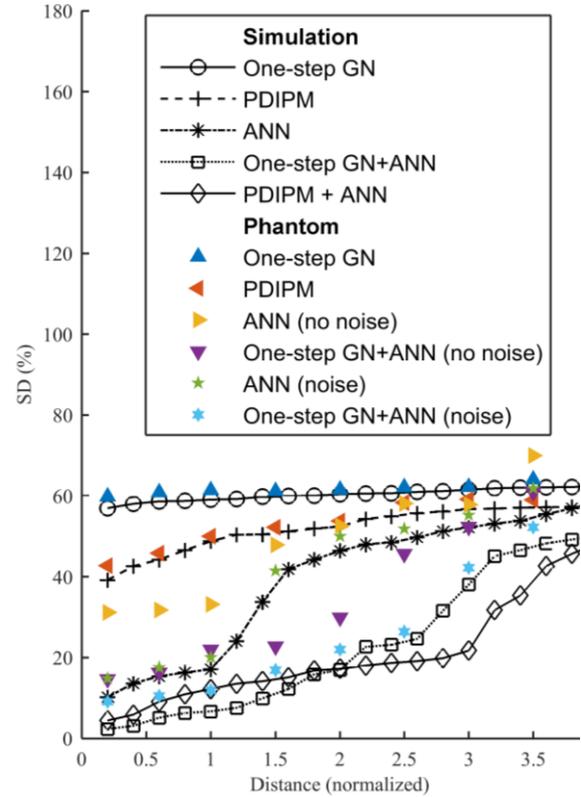

Fig 7 SD obtained for simulation and phantom data. Errors are given for different reconstruction methods for a target located at a variable distances from the probe.

NADE obtained from the simulation data are presented in Fig 6. The one-step GN method, which gives large artifacts and has a tendency to push the reconstructed object towards the outer boundary, cannot give an accurate estimate of the distance from the probe. The location of the target object can only be poorly evaluated and therefore the NADE is high. Using the one-step GN method, the NADE varies from 3.21 for a target close to the probe, to 4.70 for a target far from the probe. The PDIPM method gives an accurate estimate of the half of the profile of the target, which contributes to a lower NADE. However, this method still faces difficulties in estimating the profile of the target on the opposite side. In addition, as the target becomes further from the probe, it becomes more difficult to visually distinguish the ROI. These last two factors contribute to a higher NADE. By using this method, the NADE varies from 1.16 to 4.23. Here again, solutions based on ANN give an accurate reconstruction of the profile of the ROI at the expected location, and as a result the NADE is significantly lower. Although the ANN used directly as inverse solver gives a slightly lower NADE, the difference is too small to be visible in Fig 6. An ANN used as a direct inverse solver gives an NADE between 0.70 and 2.59, while the proposed post-processing method gives the greatest accuracy, with an error of between 0.46 and 1.97.

As for |ΔRES| and NADE, SD errors given in Fig 7, show that the proposed post-processing produces greater accuracy than commonly used methods. As the target becomes further from the probe, the presence of smoothness in the resulting images (yellow artifacts on the images Fig 4) generates a higher SD.

## 3.3 Phantom

After simulations, phantom experiments were conducted to validate the proposed method. To test the resistance of the proposed method to noisy data and previously unseen patterns, ANNs were trained with and without noise.

Reconstructions obtained from phantom data are shown in Fig 8. Similar to the figures obtained from simulation data, the linear one-step GN method gives results with large smoothness. This smoothness tends to push the reconstruction towards the outer boundary of the model and makes it difficult to estimate the distance from the probe. The fact that reconstructions from phantom data are close to reconstructions obtained from simulation data indicates that linear methods are less sensitive to noise present in the measurement data than nonlinear solvers.

Nonlinear methods are known to be more sensitive to noise. Reconstructions obtained from phantom data with the iterative PDIPM solver appear to be different from the expectation from simulation. The five different images, shown in Fig 8 ($2^{nd}$ column) differ from simulation data presented in Fig 4 ($2^{nd}$ column). The target ROI is not visible in the phantom data because of the large ringing around the probe, which makes it more difficult to see the ROI. For instance, in Fig 8 (b) (h) (n), only ringing artifacts around the probe are visible, and the target STN region is not clearly visible at the expected location.

When the ANNs are trained without including noise in the measurement data, an ANN used as an inverse solver usually performs poorly when real data are used. This can be seen in Fig 8 ($3^{rd}$ column), where the ANN was trained without including noise in the simulated voltages. Although the ANN seems to be capable of roughly estimating the location of the ROI, the resulting images contain large artifacts, and the area of the ROI appears to be significantly larger than expected, and is represented by a green ellipse on the reconstructions. This is particularly visible when the target is far from the probe.

Although ANN used as an inverse solver appears to be sensitive to noise and measurement errors, the proposed post-processing method seems to be less sensitive. In Fig 8 ($4^{th}$ column), it can be seen that the proposed method, used with an ANN trained without considering the presence of noise in the measured voltages, still gives an accurate reconstruction at a distance of three times the radius of the probe, where a simple ANN gives satisfactory quality when the distance does not exceeds twice the radius of the probe. The target ROI appears to be at the expected location and the artifacts are strongly reduced when compared to the ANN used as an inverse solver trained without any prior knowledge of the hardware system, and the average amount of noise expected from this system. This validates the idea that using a linear inverse solver before applying an ANN helps to even out the imperfections presents in the measured voltages. Although some artifacts are still visible, it remains possible to use the reconstructed images to estimate the location of the ROI.

When training ANNs with noisy data similar to the noise present in the measurement data, or reconstructions obtained with linear solver and noisy data, one can expect better performance, since the need of extrapolation to previously unseen data is reduced. In Fig 8 ($5^{th}$ column), an ANN trained with noisy voltages was used as an inverse solver. The reconstructed images show a higher resistivity within the expected target location, meaning that this method is capable of giving an accurate estimate of the target.

Similar to the ANN used as an inverse solver, it is possible to add noise in the simulated voltages before applying the linear inverse solver and train an ANN for the proposed post-processing application. In this case, the ANN is expected to perform better than the proposed method and an ANN trained without noisy data. In Fig 8 ($6^{th}$ column), it can be seen that this configuration also gives a high-quality reconstruction.

|ΔRES| and NADE were calculated and compared to the errors obtained from simulation data. |ΔRES| errors are shown in Fig 5. For the one-step GN method, the resulting |ΔRES| obtained for the phantom data are very close to the |ΔRES| obtained from simulation data with similar conditions. This result was expected from a visual inspection of the conductivity distributions shown in Fig 4 and Fig 8. When using phantom data, the resulting error is usually less than 3% higher than the error obtained from simulation with a similar ROI at the same location. This observation also confirms that the one-step GN method offers a strong robustness to inevitable imperfections present in measurement data. Due to the presence of a thresholding, |ΔRES| obtained from phantom data with the PDIPM method do not differ much from simulation, although the ROI is not clearly visible on reconstructions from phantom data. In terms of |ΔRES|, the PDIPM method still performs slightly better than the one-step GN method. An ANN used as an EIT inverse solver is known to offer very weak resistance to noisy data obtained from a physical phantom or biomedical data, if trained without assuming the presence of noise, or without any prior knowledge of the imperfections present in this

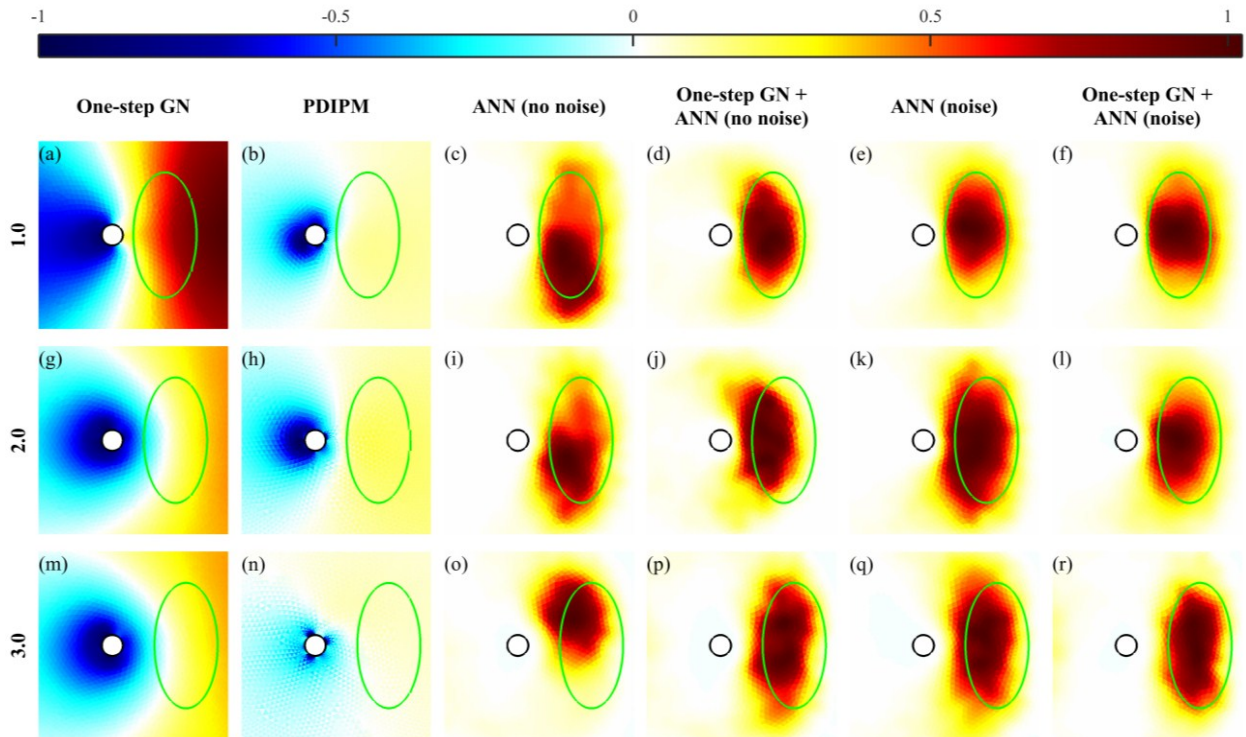

Fig 8 Comparison of EIT image reconstruction obtained from phantom data. Images were obtained with (1st column) one-step GN, (2nd column) PDIPM, (3rd column) ANN as inverse solver trained without noise, (4th column) the proposed post-processing with an ANN trained without noise, (5th column) ANN as inverse solver trained with noisy data, and (6th column) the proposed post-processing with an ANN trained with noisy data. The normalized distances between probe and target are 1, 2, and 3, respectively on 1st, 2nd, and 3rd rows. The probe is shown by a black circle and the target ROI by a green ellipse. On top, the normalized resistivity distribution.

system. In this case, the average |ΔRES| error varies from 7.92% to 28.81% for phantom data, while it does not exceed 18% on simulation data. The average |ΔRES| error obtained on phantom data with this ANN is more almost two times higher than the average |ΔRES| error obtained on simulation data. The proposed post-processing method and an ANN trained without considering the presence of noise gives a |ΔRES| error of between 4.97% and 20.79%, meaning that the proposed method offers stronger resistance to noisy data. If prior knowledge of the noise level present in the system is available, training ANNs still helps to improve the result. In this case, an ANN used as an inverse solver gives an error of between 2.79% and 25.63%. The error obtained with the proposed post-processing method and an ANN trained from noisy data varies from 0.82% to 19.61% and is at most 1.5% higher, which does not make any visible difference. In addition, as the target becomes farther from the probe, the proposed post-processing method is capable of giving a lower error than an ANN used as a direct inverse solver. With the proposed method, one can obtain satisfactory images for a distance up to 3 times the radius of the probe, as for simulations.

As expected, NADE obtained from phantom data, shown in Fig 6 with the one-step GN method, are close to NADE obtained from simulation data. With the iterative PDIPM, the NADE increases drastically as the ROI becomes far from the probe and, at an edge to edge distance of 3.5 times the radius of the probe, the resulting NADE is close to the NADE one can expect from the linear inverse solver, and is higher than the error obtained with linear solver when the target is located further away. With phantom data and the PDIPM solver, the NADE varies from 1.77 when the target is near the probe (normalized distance: 0.2), to 5.08 when the target is further away (normalized distance: 4.0). Although the NADE obtained from simulation data also increases as the target becomes farther from the probe, the increase in NADE is not as high as it is for phantom data. This confirms that the PDIPM algorithm is sensitive to errors in measurement data. As for |ΔRES| errors, an ANN used as an inverse solver trained without considering noise performs poorly. In this case, the average NADE is about 2.12 for phantom data while simulations give an average error of only 1.49 at the same locations. However, using an ANN as a post-processing tool still gives a strong resistance to noise, even when it is not considered during the training of the ANN, as shown by the resulting errors in Fig 6. In this configuration, the NADE obtained with the proposed post processing method remains very close to what one can expect from the simulation result. For each of the five data points obtained from phantom data, the NADE obtained with the proposed post-processing method, and an ANN trained without any knowledge of the noise level, is less than 0.5 higher than the NADE obtained from simulation data and the proposed method. Here again, adding noise to the simulated voltages before training the ANN helps to reduce the need for extrapolation and gives a slightly lower NADE, although the difference is insignificant for the proposed method.

Similarly, SD errors show that the proposed method gives a high accuracy. Although training the ANN by considering the presence of noise gives higher accuracy on phantom data, it can be seen that the proposed method offers strong resistance to noise. The error obtained with the proposed method and an ANN trained without including the presence of the noise, is similar to the error obtained with an ANN used as an inverse solver trained with noisy data. In both configurations (ANN used as inverse solver or post-processor), including noise in the training data results in lower errors, and higher quality.

## 3.4 Influence of inverse solvers

In this section, the proposed post-processing method is applied to images obtained with the PDIPM method in order to discuss the influence of the EIT inverse solver applied at the earliest stage of the proposed method. Images obtained this way are shown in Fig 4 (fifth column). The corresponding |ΔRES|, NADE, and SD errors are shown in Fig 5, Fig 6, and Fig 7. Fig 4 shows that that this solution also produces high quality images. Fig 5 shows that combining a nonlinear iterative method and an ANN generally gives slightly lower errors on simulation, especially when the edge to edge distance, between the target and the probe, is large. By combining the proposed post-processing method with the PDIPM solver, the NADE increases from 0.44 to 1.56. The average NADE is 0.86, which is 0.09 better than the average NADE obtained when the proposed method is combined with the one-step GN. Although the difference is small, the average |ΔRES| obtained by combining the PIDPM and an ANN is 5.46%, which is 1.58% better than the 7.04% obtained by using the one-step GN with the post-processing.

However, nonlinear iterative methods are known to be unstable when the data contain noise. This is shown by the reconstructions from phantom data in Fig 8 (2$^{nd}$ column). It appears to be difficult to see the presence of the target. To summarize, although it is possible to combine the proposed method with different EIT inverse solvers and gives good result on simulation data, the inverse solver used at the first step of the proposed method, as well as the quality of the phantom data, may strongly influence the quality of the reconstruction from phantom data. In other words, if the reconstruction method used in the early stage of the post-processing is highly sensitive to the presence of noise and other sources of errors, the proposed post-processing might not offer the high stability obtained when applying an ANN after solving the inverse problem with the one-step GN. Therefore, an accurate noise model becomes necessary to efficiently train the ANN.

## 3.5 DBS of different regions

DBS surgery is not limited to the STN region, but can be applied to different regions of the brain. Here, the proposed method is used to image a relatively small target only twice the size of the probe. Phantom experiments were conducted and the reconstructed images can be seen in Fig 9.

Similar to the case of large targets shown in Fig 8, images from the one-step GN method, shown in Fig 9 (a), (g) and (m), can give an approximate idea of the direction of the target, but cannot give a rapid estimate of the distance between the probe and the target directly from human eye.

For such a small target, the ANN may face difficulties to give a correct estimate of the conductivity distribution, especially when it is used as an inverse solver. As can be seen in Fig 9 (e), an ANN used as an inverse solver gives an accurate solution when the target is physically close to the probe. In this configuration, images obtained from the proposed method, shown in Fig 9 (d) and (f), give visual quality similar to the solution based on ANN, shown in Fig 9 (e). Fig 9 (k) and (q) both show that such an ANN, even if trained with noisy data, fails to give an accurate solution when the distance from the probe exceeds its radius.

The proposed post-processing method can image the small target with a higher degree of accuracy than the three other methods. The advantage of the proposed method becomes clear when the target is located far from the probe. In Fig 9 (r), the reconstruction shows the presence of an electrical insulator near the real location of the target, shown by a green ellipse. If the ANN is trained without considering the presence of noise, as in Fig 9 (d), (j), or (p), the proposed method does not generate large artifacts that lead to a bigger target or wrong location. Such artifacts can be seen when the ANN is used as an inverse solver, as in Fig 9 (c), (i) and (o).

Based on images shown in Fig 9, it can be said that an ANN used as an EIT inverse solver, trained using noisy voltages, gives an accurate reconstruction of the target when the distance between the probe and the target does not exceed the radius of the probe, and quickly degenerates thereafter. The proposed method gives an accurate reconstruction when the distance from the probe is more than twice larger and then slowly degenerates.

Resulting |ΔRES| errors obtained from these images are shown in Table 4. It appears that the proposed method gives a significantly smaller error. Compared to an ANN used as an inverse solver trained with noisy data, our proposed method gives a significantly lower error. When the ANN is trained with noisy data, the proposed method gives a maximal error below 15%, which represents a significant amelioration compared to the 41% from the ANN used as an inverse solver and the 69.2%

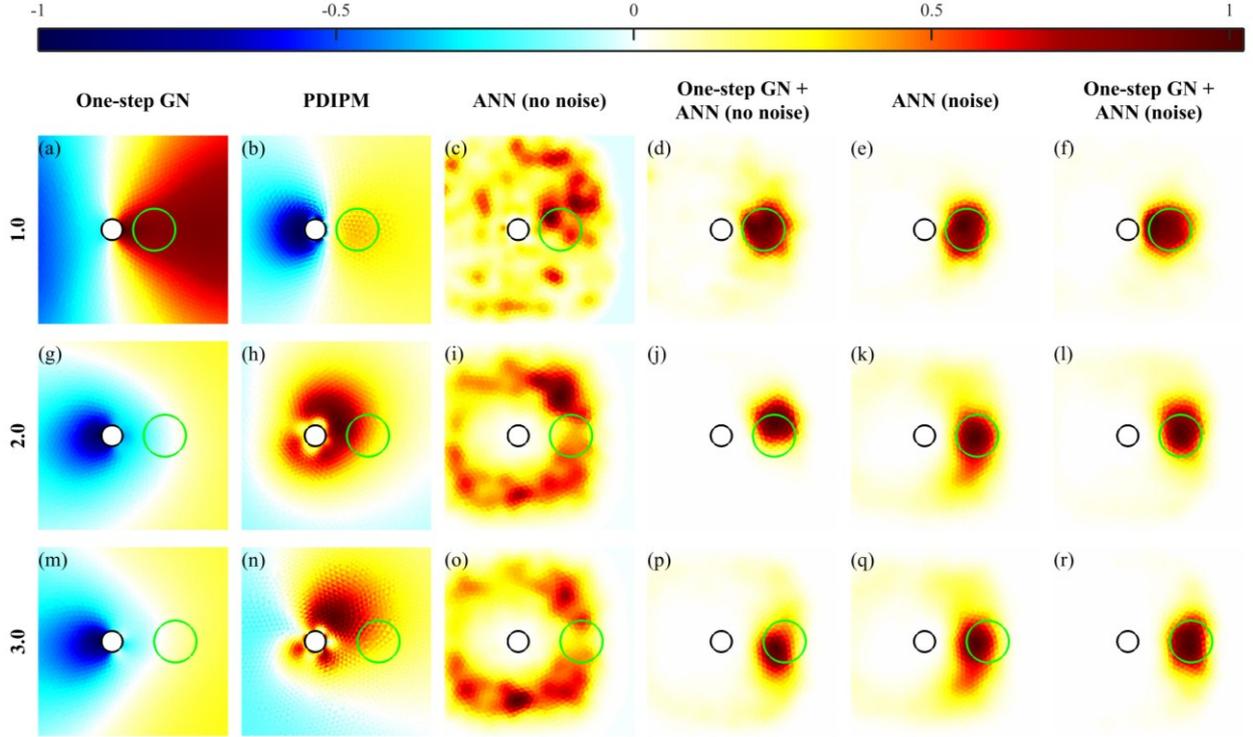

Fig 9 Comparison of EIT image reconstructions from phantom data. Images were obtained with (1st column) one-step GN, (2nd column) PDIPM, (3rd column) ANN as inverse solver trained without noise, (4th column) the proposed post-processing with an ANN trained without noise, (5th column) ANN as inverse solver trained with noisy data, and (6$^{th}$ column) the proposed post-processing with an ANN trained with noisy data. The normalized distances between probe and target (indicated on the left side) are 1, 2, 2.5, respectively on 1$^{st}$, 2$^{nd}$, and 3$^{rd}$ rows. The probe is shown by a black circle and the target ROI by a green circle. On top, the normalized resistivity distribution.

obtained from the one-step GN. Although training the ANN with noisy data results in higher quality, it can be seen that this step is not necessary with the proposed method.

Table 4 |ΔRES| errors obtained for phantom data with different reconstruction methods for a small target located at different distances from the probe. Corresponding images are shown in Fig 9

| Method \ Distance | 1.0 | 2.0 | 3.0 |
|---|---|---|---|
| One-step GN | 60.00% | 61.41% | 69.21% |
| PDIPM | 43.08% | 51.48% | 63.42% |
| ANN (training: no noise) | 40.57% | 45.20% | 50.59% |
| One-step GN + ANN (training: no noise) | 11.04% | 14.64% | 18.70% |
| ANN (training: noise) | 9.86% | 30.66% | 41.02% |
| One-step GN + ANN (training: noise) | 8.91% | 8.96% | 13.43% |

NADE errors are given in Table 5. When using the one-step GN, the PDIPM, or an ANN used as an inverse solver, large artifacts are visible on the image. Those large artifacts are responsible of a larger NADE. Here, the advantage of training the ANN used as an inverse solver with noisy data clearly appears and the average NADE error is reduced by more than half. Again, the proposed method results in greater accuracy than an ANN used as an inverse solver. The error obtained with the proposed method and an ANN trained with noisy data increases from 0.28 to 0.60. Post processing the result from one-step GN with an ANN trained without considering the presence of noise gives an error of between 0.39 and 1.00. It can be said that the proposed method, used without considering the presence of noise during the training phase, gives a similar quality that the ANN used as inverse solver trained with noise. In both case, considering the presence of noise in the training data gives a higher quality.

Table 5 NADE errors obtained for phantom data with different reconstruction methods for a small target located at different distances from the probe. Corresponding images are shown in Fig 9

| Method \ Distance | 1.0 | 2.0 | 3.0 |
|---|---|---|---|
| One-step GN | 1.82 | 2.17 | 3.05 |
| PDIPM | 1.30 | 1.76 | 2.13 |
| ANN (training: no noise) | 1.32 | 1.66 | 2.42 |
| One-step GN + ANN (training: no noise) | 0.39 | 0.89 | 1.00 |
| ANN (training: noise) | 0.32 | 0.87 | 1.00 |
| One-step GN + ANN (training: noise) | 0.28 | 0.53 | 0.60 |

Finally, the SD errors given in Table 6, confirm that the proposed method results in significantly greater accuracy, as well as a significantly greater robustness to noise. If the ANN is trained without considering the presence of noise in the training data, the resulting error is at least 86%, while the proposed method gives an error of at most 58%. Including the presence of noise in the training data helps improve the quality of reconstruction. In this case, the proposed method gives an error of at most 44%, where the ANN used as an inverse solver gives an error of between 49% and 75%.

Table 6 SD errors obtained for phantom data with different reconstruction methods for a small target located at different distances from the probe. Corresponding images are shown in Fig 9

| Method \ Distance | 1.0 | 2.0 | 3.0 |
|---|---|---|---|
| One-step GN | 92.54% | 94.19% | 93.89% |
| PDIPM | 92.97% | 92.26% | 94.51% |
| ANN (training: no noise) | 86.42% | 91.42% | 91.33% |
| One-step GN + ANN (training: no noise) | 49.32% | 51.67% | 58.03% |
| ANN (training: noise) | 49.04% | 64.26% | 75.06% |
| One-step GN + ANN (training: noise) | 43.18% | 43.96% | 44.08% |

## 3.6 Computation resources

For each different method, CPU time and required memory to solve the EIT inverse problem were calculated and the results are presented in Table 7.

Table 7 CPU time and required memory for solving the EIT inverse problem with different methods

| | one-step GN | PDIPM | ANN | one-step GN+ANN |
|---|---|---|---|---|
| Time (ms) | 32 | 2632 | 126 | 283 |
| Memory (Gb) | <0.1 | 0.8 | 0.3 | 0.8 |

It appears that the fastest method remains the linear one-step GN, which consists of a single matrix product. Computing this reconstruction matrix is a costly task that requires 1.1 Gb of RAM and 5.1 seconds on the machine described in this paper. However, since it does not depend on the measurements, this step can be done prior to data acquisition, which leads to a faster reconstruction. If the reconstruction matrix is known, this method estimates the conductivity distribution within 32 ms, and requires a very small amount of memory. The nonlinear method based on ANN used as an inverse solver is approximately 4 times slower than the one-step GN and can give a nonlinear estimate of the conductivity distribution in 126 ms, requiring 0.3 Gb of RAM in this implementation. The proposed method, which combines the one-step GN method and ANN gives a solution in 283 ms and requires up to 0.8 Gb of RAM. This larger amount is due to the size of the ANN used in this method. In our proposed method, the number of inputs is equivalent to the number of nodes in the FEM, which is higher than the number of available measurements. For this reason, the ANN used in the proposed method contains more input neurons than the ANN used at an inverse solver, which then leads to a longer computation time on a PC. Although the proposed method is significantly longer than one-step GN or the ANN used as an inverse solver, it still can be considered as real-time, since in real DBS experiments the probe moves slowly. Finally, the iterative PDIPM method is more than 9 times slower than the proposed post-processing method. As it requires more than 2.6 seconds estimating a solution to the inverse problem, this approach is too time consuming to be regarded as a real-time method.

## 4 DISCUSSION

This paper constitutes both a feasibility study and a novel method of optimizing the solution to the EIT inverse problem in an open-domain. Practical applications of open-domain EIT are known to be more complicated than closed domain EIT problems, especially because of the presence of noise, which is expected to be relatively high in real DBS applications, caused by the electrical current naturally emitted by the brain. For this reason, direct nonlinear methods and nonlinear iterative methods are very likely to be badly influenced by the noise in the data, generate poor approximation, and therefore are not applicable in practice.

The proposed reconstruction method, by combining linear and nonlinear algorithms, has the advantages of both and minimal drawbacks. Not only does this method offers strong robustness against noise, but also provides high quality images of the electrical conductivity distribution. When using the proposed method, the ANN is capable of generating an accurate estimate of the conductivity distribution within the phantom, even if trained without the presence of noise. This last observation demonstrates that, compared to the solution based on ANN applied directly on the measured data, which requires including the presence of noise in the training data to obtain a satisfactory image, and therefore having an a priori knowledge of this specific noise level, the proposed method offers a strong resistance to noisy data.

## 5 CONCLUSIONS

With this strong robustness to noisy data, the proposed method is less likely to fail on animal studies or human experiments, in which cases the modelling effort required to generate training data and then train an ANN is significantly more important than for phantom experiments. In addition, one can always argue that the physiological artefacts are not perfectly represented in the training dataset. However, it is expected that the proposed post-processing method will give high quality real-time images even in case the ANN was trained with approximate data. This idea can later facilitate the training of the ANN for in-vivo reconstruction.

# ACKNOWLEDGMENTS

The author would like to thank Chii-Chew Hong for designing the hardware system, the EIDORS project for their open source library, and the Ministry Of Science and Technology (MOST) of Taiwan for funding this research.

# REFERENCES

Abascal, J.-F.P.J., Arridge, S.R., Bayford, R.H., Holder, D.S., 2008. Comparison of methods for optimal choice of the regularization parameter for linear electrical impedance tomography of brain function. Physiol. Meas. 29, 1319–34. doi:10.1088/0967-3334/29/11/007

Adler, A., Lionheart, W.R.B., 2006. Uses and abuses of EIDORS: an extensible software base for EIT. Physiol. Meas. 27, S25–S42. doi:10.1088/0967-3334/27/5/S03

Baizabal Carvallo, J.F., Mostile, G., Almaguer, M., Davidson, A., Simpson, R., Jankovic, J., 2012. Deep Brain Stimulation Hardware Complications in Patients with Movement Disorders: Risk Factors and Clinical Correlations. Stereotact. Funct. Neurosurg. 90, 300–306. doi:10.1159/000338222

Bayford, R.H., 2006. Bioimpedance tomography (electrical impedance tomography). Annu. Rev. Biomed. Eng. 8, 63–91. doi:10.1146/annurev.bioeng.8.061505.095716

Boon, P., Vonck, K., De Herdt, V., Van Dycke, A., Goethals, M., Goossens, L., Van Zandijcke, M., De Smedt, T., Dewaele, I., Achten, R., Wadman, W., Dewaele, F., Caemaert, J., Van Roost, D., 2007. Deep brain stimulation in patients with refractory temporal lobe epilepsy. Epilepsia 48, 1551–60. doi:10.1111/j.1528-1167.2007.01005.x

Borsic, A., Graham, B.M., Adler, A., Lionheart, W.R.B., 2010. In vivo impedance imaging with total variation regularization. IEEE Trans. Med. Imaging 29, 44–54. doi:10.1109/TMI.2009.2022540

Borsic, A., Halter, R., Wan, Y., Hartov, A., Paulsen, K.D., 2010. Electrical impedance tomography reconstruction for three-dimensional imaging of the prostate. Physiol. Meas. 31, S1-16. doi:10.1088/0967-3334/31/8/S01

Brown, B.H., 2003. Electrical impedance tomography (EIT): a review. J. Med. Eng. Technol. 27, 97–108. doi:10.1080/0309190021000059687

Burn, D.J., Tröster, A.I., 2004. Neuropsychiatric complications of medical and surgical therapies for Parkinson's disease. J. Geriatr. Psychiatry Neurol. 17, 172–180. doi:10.1177/0891988704267466

Chaturvedi, A., 2012. Development of accurate computational models for patient-specific deep brain stimulation. Case Western Reserve University.

Chen, Q., Konrad, A., 1997. A review of finite element open boundary techniques for static and quasi-static electromagnetic field problems. IEEE Trans. Magn. 33, 663–676. doi:10.1109/20.560095

Cheney, M., Isaacson, D., Newell, J.C., 1999. Electrical Impedance Tomography. SIAM Rev. 41, 85–101. doi:10.1137/S0036144598333613

Fiegele, T., Feuchtner, G., Sohm, F., Bauer, R., Anton, J.V., Gotwald, T., Twerdy, K., Eisner, W., 2008. Accuracy of stereotactic electrode placement in deep brain stimulation by intraoperative computed tomography. Parkinsonism Relat. Disord. 14, 595–599. doi:10.1016/j.parkreldis.2008.01.008

Holder, D.S., 2005. Electrical Impedance Tomography: Methods, History and Applications. Institute of Physics Publishing, London, U.K.

Hu Hen, Y., Hwang, J.-N., 2001. Handbook of neural network signal processing, 1st ed. CRC Press, London, U.K.

Kuen, J., Woo, E.J., Seo, J.K., 2009. Multi-frequency time-difference complex conductivity imaging of canine and human lungs using the KHU Mark1 EIT system. Physiol. Meas. 30, S149-64. doi:10.1088/0967-3334/30/6/S10

Laxton, A.W., Tang-Wai, D.F., McAndrews, M.P., Zumsteg, D., Wennberg, R., Keren, R., Wherrett, J., Naglie, G., Hamani, C., Smith, G.S., Lozano, A.M., 2010. A phase I trial of deep brain stimulation of memory circuits in Alzheimer's disease. Ann. Neurol. 68, 521–534. doi:10.1002/ana.22089

Leonhardt, S., Lachmann, B., 2012. Electrical impedance tomography: the holy grail of ventilation and perfusion monitoring? Intensive Care Med. 38, 1917–29. doi:10.1007/s00134-012-2684-z

Leshno, M., Lin, V., Pinkus, A., Schocken, S., 1993. Multilayer Feedforward Networks with a Non-Polynomial Activation Function Can Approximate Any Function. Neural Networks 6, 861–867.

Liker, M. a., Won, D.S., Rao, V.Y., Hua, S.E., 2008. Deep Brain Stimulation: An Evolving Technology. Proc. IEEE 96, 1129–1141. doi:10.1109/JPROC.2008.922559

Lozano, A.M., 2001. Deep brain stimulation for Parkinson's disease. Parkinsonism Relat. Disord. 7, 199–203. doi:10.1016/S1353-8020(00)00057-2

Martin, S., Choi, C.T.M., 2017. A Post-Processing Method for Three-Dimensional Electrical Impedance Tomography. Sci. Rep. 7, 7212. doi:10.1038/s41598-017-07727-2

Martin, S., Choi, C.T.M., 2016. Nonlinear Electrical Impedance Tomography Reconstruction Using Artificial Neural Networks and Particle Swarm Optimization. IEEE Trans. Magn. 52, 1–4. doi:10.1109/TMAG.2015.2488901

McIlwain, H., 1951. Metabolic response in vitro to electrical stimulation of sections of mammalian brain. Biochem. J. 49, 382–93.

McIntyre, C.C., Chaturvedi, A., Shamir, R.R., Lempka, S.F., 2015. Engineering the Next Generation of Clinical Deep Brain Stimulation Technology. Brain Stimul. 8, 21–26. doi:10.1016/j.brs.2014.07.039

Miao, L., Ma, Y., Wang, J., 2014. ROI-Based Image Reconstruction of Electrical Impedance Tomography Used to Detect Regional Conductivity Variation. IEEE Trans. Instrum. Meas. 63, 2903–2910. doi:10.1109/TIM.2014.2326765

Perlmutter, J.S., Mink, J.W., 2006. Deep brain stimulation. Annu. Rev. Neurosci. 29, 229–257. doi:10.1146/annurev.neuro.29.051605.112824

Tu, C.-Y., 2015. Development of High Resolution Electrical Impedance Tomography System. National Chiao Tung University.

Vayssiere, N., Hemm, S., Zanca, M., Picot, M.C., Bonafe, A., Cif, L., Frerebeau, P., Coubes, P., 2000. Magnetic resonance imaging stereotactic target localization for deep brain stimulation in dystonic children. J. Neurosurg. 93, 784–790. doi:10.3171/jns.2000.93.5.0784

Wang, P., Li, H., Xie, L., Sun, Y., 2009. The Implementation of FEM and RBF Neural Network in EIT, in: Second International Conference on Intelligent Networks and Intelligent Systems. Ieee, pp. 66–69. doi:10.1109/ICINIS.2009.26

Winkler, D., 2005. The first evaluation of brain shift during functional neurosurgery by deformation field analysis. J. Neurol. Neurosurg. Psychiatry 76, 1161–1163. doi:10.1136/jnnp.2004.047373

Xu, G., Wu, H., Yang, S., Liu, S., Li, Y., Yang, Q., Yan, W., Wang, M., 2005. 3-D electrical impedance tomography forward problem with finite element method. IEEE Trans. Magn. 41, 1832–1835. doi:10.1109/TMAG.2005.846503

Zhou, B., Xu, C., Yang, D., Wang, S., Wu, X., 2007. Nonlinear image reconstruction using a GA-ECT technique in electrical capacitance tomography. Flow Meas. Instrum. 18, 285–294. doi:10.1016/j.flowmeasinst.2007.06.011